\documentclass[conference]{IEEEtran}
\IEEEoverridecommandlockouts

\usepackage{cite}
\usepackage{amsmath,amssymb,amsfonts}
\usepackage{mathtools}
\usepackage{bm}
\usepackage{hyperref}
\usepackage{algorithm}
\usepackage{algpseudocode}
\usepackage{graphicx}
\usepackage{textcomp}
\usepackage{xcolor}
\usepackage{braket}
\usepackage{qcircuit}
\usepackage{siunitx}   
\usepackage{booktabs}  
\usepackage{stfloats} 
\usepackage{lmodern}     
\usepackage[T1]{fontenc}  

\usepackage{tikz}
\usetikzlibrary{arrows.meta,positioning,fit,shapes.misc}

\usepackage{hyperref}
\hypersetup{
    colorlinks=true,
    linkcolor=blue,
    urlcolor=cyan,
}

\usepackage{atbegshi}
\usepackage{eso-pic}
\AtBeginShipout{%
  \ifnum\value{page}=4
\global\setbox\AtBeginShipoutBox=\vbox{\vspace{2pt}\unvbox\AtBeginShipoutBox}%
  \fi
}
\usetikzlibrary{positioning, calc, shapes.geometric, arrows.meta}

\def\BibTeX{{\rm B\kern-.05em{\sc i\kern-.025em b}\kern-.08em
    T\kern-.1667em\lower.7ex\hbox{E}\kern-.125emX}}

\begin{document}

\title{Quantum Adaptive Self-Attention for Financial Rebalancing: An Empirical Study on Automated Market Makers in Decentralized Finance

}

\author{
\IEEEauthorblockN{Chi-Sheng Chen}
\IEEEauthorblockA{\textit{Omnis Labs} \\
Cambridge, USA \\
m50816m50816@gmail.com}
\and
\IEEEauthorblockN{Aidan Hung-Wen Tsai}
\IEEEauthorblockA{\textit{Omnis Labs} \\
 New York, USA \\
aidan@omnis.farm}
}

\maketitle

\begin{abstract}
We formulate automated market maker (AMM) \emph{rebalancing} as a binary detection problem and study a hybrid quantum--classical self-attention block, \textbf{Quantum Adaptive Self-Attention (QASA)}. QASA constructs quantum queries/keys/values via variational quantum circuits (VQCs) and applies standard softmax attention over Pauli-$Z$ expectation vectors, yielding a drop-in attention module for financial time-series decision making. Using daily data for \textbf{BTCUSDC} over \textbf{Jan-2024--Jan-2025} with a 70/15/15 time-series split, we compare QASA against classical ensembles, a transformer, and pure quantum baselines under Return, Sharpe, and Max Drawdown. The \textbf{QASA-Sequence} variant attains the \emph{best single-model risk-adjusted performance} (\textbf{13.99\%} return; \textbf{Sharpe 1.76}), while hybrid models average \textbf{11.2\%} return (vs.\ 9.8\% classical; 4.4\% pure quantum), indicating a favorable performance--stability--cost trade-off. 
\end{abstract}


\section{Introduction}
Decentralized finance (DeFi) has grown rapidly, with automated market makers (AMMs) becoming core primitives for on-chain liquidity provisioning \cite{Adams2021UniswapV3}. AMMs implement constant-function designs (CFMMs) whose pricing and value properties have been formalized in recent theory, clarifying when pool prices are informative and how payoff replication arises \cite{Angeris2020CFMM,Angeris2021RepMM}. For liquidity providers (LPs), the practical challenge is \emph{when and how to rebalance} under volatile, regime-switching markets. Empirical and theoretical studies further decompose LP P\&L into market risk and \emph{loss-versus-rebalancing (LVR)}, highlighting adverse selection and fee interplay \cite{Milionis2022LVR}, and show that concentrated-liquidity designs (e.g., Uniswap v3) require active management to maintain performance \cite{Heimbach2022V3,Neuder2021SLP}.

Existing approaches to rebalancing span (i) transparent heuristics (fixed intervals, volatility triggers) and (ii) predictive machine learning. However, financial time series are noisy, non-stationary, and often governed by latent regimes \cite{Hamilton1989,Tsay2010}, making generalization fragile. While attention-based architectures such as Transformers \cite{Vaswani2017} have delivered strong results in market microstructure and forecasting—e.g., limit-order-book (LOB) prediction and multi-horizon forecasting \cite{Zhang2019DeepLOB,Wallbridge2020TransLOB,Lim2021TFT}—their data hunger and overfitting risk can limit robustness in small-sample or rapidly shifting regimes typical for on-chain markets.

To address these challenges, we build on the \emph{Quantum Adaptive Self-Attention (QASA)} block introduced in prior work \cite{Chen2025QASA} and tailor it to DeFi: compact descriptors (price changes, realized volatility, flow imbalance) are embedded into qubits and processed via VQCs; the measured expectations parameterize queries, keys, and values in a classical attention head \cite{Cherrat2024QViT}. From the feature-space viewpoint, quantum encodings act as (potentially) high-dimensional nonlinear maps \cite{Havlicek2019,Schuld2019FeatureHilbert,Schuld2021Kernel}, providing expressive interactions and inductive biases helpful for regime detection on noisy data, while remaining NISQ-friendly \cite{Preskill2018, chen2025benchmarking}. Early quantum–attention explorations (e.g., quantum Transformers and quantum self-attention) suggest architectural viability across modalities \cite{Cherrat2024QViT}; in finance, quantum amplitude-estimation pipelines already deliver algorithmic speedups for option pricing \cite{Stamatopoulos2020QAE}.

\paragraph{Problem setup.}
We cast rebalancing as a detection task. Labels are defined by thresholding deviations from a moving-average anchor:
\begin{equation}
y_t=\mathbb{I}\!\left(\frac{P_t}{\mathrm{MA}_{20}(P_t)}-1>\tau_{\text{rebalance}}\right), \quad \tau_{\text{rebalance}}=0.02,
\end{equation}
where $P_t$ is the spot price and $\mathrm{MA}_{20}$ a 20-period moving average. We evaluate strategies with standard trading metrics (Return, Sharpe, Max Drawdown) under identical fee and cooldown settings to ensure fair comparisons.

\paragraph{Contributions.}
(i) We present, to the best of our knowledge, the \emph{first application of prior QASA-style quantum self-attention} to DeFi/AMM rebalancing, adapting the block introduced in earlier work \cite{Chen2025QASA}—and related quantum self-attention designs \cite{Cherrat2024QViT}—to a detection-oriented rebalancing pipeline with DeFi-specific encodings (fees, ticks, cooldowns).
(ii) We tailor QASA to AMM data via two practical instantiations: \emph{QASA-Sequence} (short raw price windows) and \emph{QASA-Hybrid} (raw windows plus engineered microstructure features), where Q/K/V are generated by variational quantum circuits and consumed by a classical attention head.
(iii) In a controlled 2024 backtest on three liquid crypto pairs with strict time-series splits and identical fee/cooldown settings, \textbf{QASA-Sequence} attains the best Sharpe (1.76) and return (13.99\%), while hybrid variants outperform pure-quantum baselines on average.

\section{Related Work}
\paragraph{AMMs and concentrated liquidity.}
Automated market makers (AMMs) based on constant‑function designs (CFMMs) have been analyzed from both protocol and economic perspectives.
On the protocol side, Uniswap v2/v3 technical whitepapers specify the hardened oracle and concentrated‑liquidity mechanisms that motivate rebalancing policy design under fees and tick ranges \cite{Adams2020UniswapV2,Adams2021UniswapV3}.
Foundational analyses of CFMMs (e.g., price oracles, convexity, replication) formalize when on‑chain prices are informative and how CFMM payoff surfaces relate to desired monotone payoffs \cite{Angeris2020CFMM,Angeris2021RepMM}.
From the LP perspective, Milionis et al.\ decompose liquidity‑provision P\&L into market‑risk and \emph{loss‑versus‑rebalancing (LVR)} components, clarifying the role of volatility, depth and fees in profitability \cite{Milionis2022LVR}.
For Uniswap v3 specifically, strategic/tactical LP studies and empirical audits highlight the need for \emph{active} management and regime awareness \cite{Neuder2021SLP,Heimbach2022V3}.

\paragraph{Learning for market regimes and volatility.}
Deep sequence models are widely used in financial time series.
For high‑frequency microstructure, CNN/LSTM hybrids (DeepLOB) established strong baselines on limit‑order‑book (LOB) data \cite{Zhang2018DeepLOB}, and more recent Transformer‑style architectures show further gains on LOB classification and forecasting \cite{Wallbridge2020TransLOB}.
For lower‑frequency multi‑horizon forecasting, the Temporal Fusion Transformer (TFT) offers competitive accuracy with interpretability and has been applied to realized‑volatility prediction in turbulent markets \cite{Lim2021TFT,Frank2023TFTVol}.

\paragraph{Quantum ML and quantum attention.}
Quantum machine learning (QML) can be interpreted through feature‑space/kernel lenses \cite{Schuld2019FeatureHilbert,Schuld2021Kernel, chen2025quantumtxt} and has been explored in finance via amplitude‑estimation–based pricing/risk algorithms \cite{Stamatopoulos2020QAE} and portfolio optimization.
Closer to attention mechanisms, quantum Transformer variants and quantum self‑attention modules have been proposed and evaluated on vision and NLP tasks \cite{Cherrat2024QViT,Li2022QSANN}.
Departing from prior fully‑quantum attention layers, our \textbf{QASA} combines compact feature encodings with variational quantum circuits (VQCs) to produce query/key/value statistics consumed by a \emph{classical} attention head, aiming for NISQ‑friendly training while preserving classical throughput \cite{Chen2025QASA}.

\begin{figure}
    \centering
    \includegraphics[width=1\linewidth]{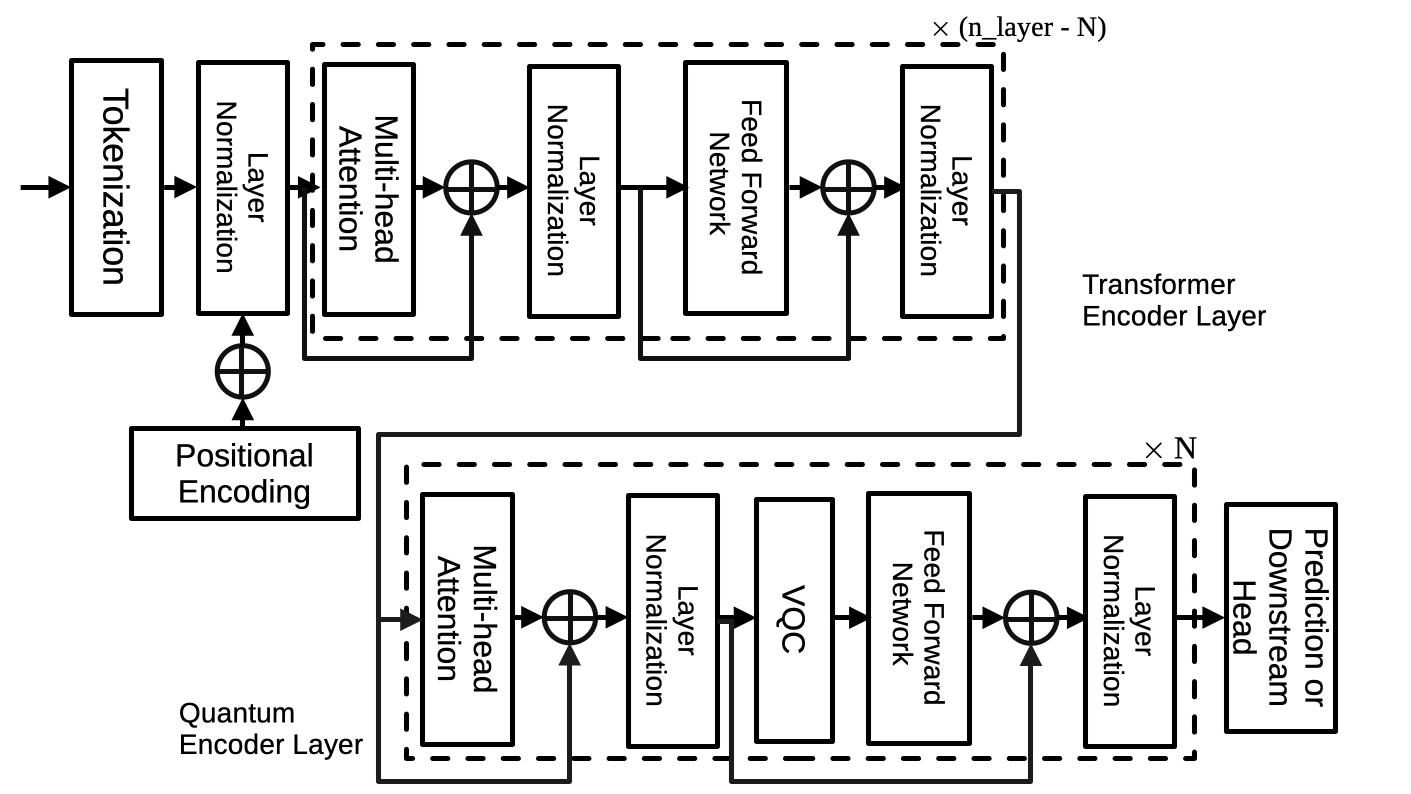}
    \caption{QASA model.}
    \label{fig:qasa}
\end{figure}

\section{Methodology}
\textbf{Task.} Rebalancing as detection via the MA-anchored threshold in (1). We also consider alternative labels (concentrated-liquidity band and price-change thresholds) in ablations. 

\textbf{Data.} Daily candles for BTCUSDC from \emph{2024-01-01} to \emph{2025-01-01} (about 252 trading days); \emph{70/15/15} time-series split (train/val/test). Five-run repeats quantify uncertainty. 

\paragraph{Features \& quantum mapping.}
\textbf{Notation.}
Let $t\in\mathbb{Z}$ index bars (e.g., 5-min). Denote mid/close price $P_t$, open $O_t$, high $H_t$, low $L_t$, close $C_t$, and volume $V_t$.
Define log-returns $r_t := \log(P_t)-\log(P_{t-1})$ and the lag operator $\mathcal{L}^\ell x_t := x_{t-\ell}$.

\medskip
\noindent\textbf{Moving averages and volatility.}
For window $n\ge 1$,
\begin{align}
\mathrm{MA}_n(P_t) &:= \frac{1}{n}\sum_{j=0}^{n-1} P_{t-j},\\
\mathrm{EMA}_n(x_t) &:= \alpha_n x_t + (1-\alpha_n)\,\mathrm{EMA}_n(x_{t-1}),\quad \alpha_n=\tfrac{2}{n+1},\\
\hat\sigma_{n,t}^2 &:= \frac{1}{n-1}\sum_{j=1}^{n}\big(r_{t+1-j}-\bar r_{n,t}\big)^2,\quad \bar r_{n,t}:=\frac{1}{n}\sum_{j=1}^{n} r_{t+1-j},\\
\hat\sigma^{\mathrm{EWMA}}_{t}{}^{2} &:= (1-\lambda)\,r_t^2 + \lambda\,\hat\sigma^{\mathrm{EWMA}}_{t-1}{}^{2},\quad \lambda\in(0,1).
\end{align}

\medskip
\noindent\textbf{Price/MA ratios and momentum.}
For windows $k,n$,
\begin{align}
\text{(MA ratio)}\quad \rho_{n}(t) &:= \frac{P_t}{\mathrm{MA}_n(P_t)} - 1,\\
\text{(k-step momentum)}\quad m_{k}(t) &:= \sum_{j=1}^{k} r_{t+1-j} \;=\; \log\!\frac{P_t}{P_{t-k}}.
\end{align}


\medskip
\noindent\textbf{Bollinger bands and ATR.}
For $(n_{\mathrm{BB}},k_{\mathrm{BB}})$,
\begin{align}
\mu_t &:= \mathrm{MA}_{n_{\mathrm{BB}}}(P_t),\quad
s_t := \sqrt{\frac{1}{n_{\mathrm{BB}}-1}\sum_{j=0}^{n_{\mathrm{BB}}-1}\big(P_{t-j}-\mu_t\big)^2},\\
U_t &:= \mu_t + k_{\mathrm{BB}} s_t,\quad L_t := \mu_t - k_{\mathrm{BB}} s_t,\\
\%b_t &:= \frac{P_t - L_t}{U_t - L_t}\in[0,1],\qquad
z^{\mathrm{BB}}_t := \frac{P_t-\mu_t}{k_{\mathrm{BB}} s_t}.
\end{align}
True range and ATR over $n_{\mathrm{ATR}}$:
\begin{align}
\mathrm{TR}_t &:= \max\{H_t-L_t,\,|H_t-C_{t-1}|,\,|L_t-C_{t-1}|\},\\
\mathrm{ATR}_t &:= \mathrm{EMA}_{n_{\mathrm{ATR}}}(\mathrm{TR}_t),\qquad
a^{\mathrm{rel}}_t := \frac{\mathrm{ATR}_t}{P_t}.
\end{align}

\medskip
\noindent\textbf{Volume and microstructure proxies.}
\begin{align}
\text{(volume ratio)}\quad \upsilon_{n}(t) &:= \frac{V_t}{\mathrm{MA}_n(V_t)}-1,\\
\text{(signed volume)}\quad S_t &:= \mathrm{sgn}(r_t)\,V_t,\quad \\
\tilde\upsilon_{n}(t) := \frac{S_t}{\mathrm{MA}_n(|S_t|)+\varepsilon},\\
\text{(Amihud illiquidity)}\quad \mathcal{I}_{n}(t) &:= \frac{1}{n}\sum_{j=0}^{n-1}\frac{|r_{t-j}|}{V_{t-j}+\varepsilon}.
\end{align}

\medskip
\noindent\textbf{RSI, MACD, Lags and interactions.}
For a base feature set $\mathcal{F}=\{x_t^{(1)},\dots,x_t^{(d)}\}$ and lag set $\mathcal{L}\subset\mathbb{N}$,
\begin{align}
\text{(lag-augmented)}\quad \mathbf{x}_t &:= \big[x_t^{(1)},\dots,x_t^{(d)},\{\mathcal{L}^\ell x_t^{(j)}\}_{j,\ell\in\mathcal{F}\times\mathcal{L}}\big],\\
\text{(interactions)}\quad \mathcal{I}_{\mathrm{pair}}(t) &:= \{x_t^{(i)}x_t^{(j)}:\,(i,j)\in\mathcal{S}\subset\{1,\dots,d\}^2\}.
\end{align}

\medskip
\noindent\textbf{From features to six scalars (for qubits).}
We aggregate engineered features into six continuous channels
\[
\mathbf{s}_t=\big[s^{(1)}_t,\dots,s^{(6)}_t\big]
\]
as
\begin{align}
s^{(1)}_t &:= m_{k_m}(t)\quad\text{(momentum)},\\
s^{(2)}_t &:= \rho_{n_\rho}(t)\quad\text{(price/MA ratio)},\\
s^{(3)}_t &:= \nu_t:=\log\!\frac{\hat\sigma^{\mathrm{EWMA}}_{t}}{\hat\sigma_{n_{\mathrm{LR}},t}}\quad\text{(volatility regime)},\\
\underbrace{s^{(4a)}_t}_{\mathrm{RSI}}\!&:= \frac{\mathrm{RSI}_t}{100},\qquad
\underbrace{s^{(4b)}_t}_{\mathrm{MACD}}\!:= \mathrm{Hist}_t,\\
s^{(5)}_t &:= \upsilon_{n_v}(t)\quad\text{(volume ratio)},\\
\underbrace{s^{(6a)}_t}_{\mathrm{BB\ position}}\!&:= \%b_t,\qquad
\underbrace{s^{(6b)}_t}_{\mathrm{ATR\ rel}}\!:= a^{\mathrm{rel}}_t.
\end{align}
Here $n_{\mathrm{LR}}$ is a long-run window (e.g., days) for baseline volatility.

\medskip
\noindent\textbf{Angle encoding (min–max, train-split only).}
For any scalar channel $z_t$, define
\begin{align}
\mathrm{mm}(z_t;a,b) &:= \mathrm{clip}\!\left(\frac{z_t-a}{\,b-a\,},\,0,\,1\right),\qquad
a:=\min_{\tau\in\mathcal{T}_{\mathrm{train}}}z_\tau,\; b:=\max_{\tau\in\mathcal{T}_{\mathrm{train}}}z_\tau,\\
\theta[z_t] &:= 2\pi\,\mathrm{mm}(z_t;a,b).
\end{align}
The 6-qubit angle map uses one rotation per scalar channel, with two-axis encodings for composite channels:
\begin{align}
&\text{Initialize}\quad |\psi_{\mathrm{in}}\rangle = |0\rangle^{\otimes 6}.\\
&\text{Qubit 1 (momentum)}:\quad R_y\!\big(\theta[s^{(1)}_t]\big),\\
&\text{Qubit 2 (MA ratio)}:\quad R_y\!\big(\theta[s^{(2)}_t]\big),\\
&\text{Qubit 3 (vol regime)}:\quad R_y\!\big(\theta[s^{(3)}_t]\big),\\
&\text{Qubit 4 (RSI/MACD)}:\quad R_y\!\big(\theta[s^{(4a)}_t]\big)\,R_z\!\big(\theta[s^{(4b)}_t]\big),\\
&\text{Qubit 5 (volume ratio)}:\quad R_y\!\big(\theta[s^{(5)}_t]\big),\\
&\text{Qubit 6 (BB/ATR)}:\quad R_y\!\big(\theta[s^{(6a)}_t]\big)\,R_z\!\big(\theta[s^{(6b)}_t]\big).
\end{align}
This yields a feature-encoded state $|\psi_t\rangle := U_{\mathrm{enc}}(t)\,|\psi_{\mathrm{in}}\rangle$; a variational circuit $U_{\mathrm{VQC}}(\boldsymbol{\phi})$ then acts on $|\psi_t\rangle$, and measured expectations $\langle Z_i\rangle$ feed the classical attention head as Q/K/V.

\textbf{Models.} 
\paragraph{QASA mechanism}
Given a token (or feature) vector $\mathbf{x}_t\in\mathbb{R}^d$, QASA uses three parameterized variational quantum circuits (VQCs) to produce query, key, and value:
\begin{align}
Q_t &= \mathrm{VQC}_q(\mathbf{x}_t),\quad
K_t = \mathrm{VQC}_k(\mathbf{x}_t),\quad
V_t = \mathrm{VQC}_v(\mathbf{x}_t).
\label{eq:qasa_qkv}
\end{align}

\textit{State preparation.} For sequence inputs, we use amplitude encoding:
\begin{equation}
\lvert \psi_t \rangle \;=\; \frac{1}{\lVert \mathbf{x}_t \rVert_2}\sum_{i=1}^{d} x_{t,i}\,\lvert i\rangle .
\label{eq:qasa_amp}
\end{equation}
For engineered features (hybrid variant), we may use angle encoding:
\begin{equation}
\theta_i \;=\; \frac{x_i - x_{\min}}{x_{\max}-x_{\min}}\cdot 2\pi ,
\label{eq:qasa_angle}
\end{equation}
and feed $\{\theta_i\}$ into the circuit as single-qubit parameterized rotations. 

\textit{Variational circuit.} On $n=\lceil\log_2 d\rceil$ qubits with $L$ layers, a typical VQC block is
\begin{equation}
U(\boldsymbol{\theta}) \;=\; \prod_{\ell=1}^{L}\!\left(\;\prod_{i=1}^{n} R_Y(\theta_{\ell,i})\right)
\left(\prod_{i=1}^{n-1}\mathrm{CNOT}_{i,i+1}\right),
\label{eq:qasa_unitary}
\end{equation}
where $R_Y(\cdot)$ are parameterized rotations and CNOTs provide entanglement.

\textit{Measurement.} Each VQC outputs an $n$-dimensional expectation vector by Pauli-$Z$ measurements:
\begin{equation}
\mathrm{VQC}(\mathbf{x}_t) \;=\; \big(\langle Z_1\rangle,\,\ldots,\,\langle Z_n\rangle\big),\qquad
\langle Z_i\rangle \;=\; \langle \psi_t \lvert U^{\dagger}(\boldsymbol{\theta})\, Z_i\, U(\boldsymbol{\theta}) \rvert \psi_t\rangle .
\label{eq:qasa_measure}
\end{equation}

\textit{Attention.} Given the stacked matrices $Q, K, V$ over a window (or mini-batch), QASA applies classical scaled dot-product attention:
\begin{equation}
\mathrm{Attn}(Q,K,V)\;=\; \operatorname{softmax}\!\left(\frac{QK^{\top}}{\sqrt{d_v}}\right)V,
\label{eq:qasa_attn}
\end{equation}
with $d_v$ the value dimension. The attention output is then fed to a shallow head for prediction.

\emph{QASA-Sequence}: price windows $\rightarrow$ VQC-generated Q/K/V $\rightarrow$ attention $\rightarrow$ classifier. 
\emph{QASA-Hybrid}: engineered features $\rightarrow$ VQC Q/K/V $\rightarrow$ attention. 
Baselines: RF, GB, Logistic Regression, Transformer, and pure quantum (VQE, QNN, QSVM). 

\textbf{QASA block.} For token $\mathbf{x}_t$, 
\begin{align}
\begin{split}
Q_t&=\mathrm{VQC}_q(\mathbf{x}_t),\quad K_t=\mathrm{VQC}_k(\mathbf{x}_t),\quad  \\ V_t=\mathrm{VQC}_v(\mathbf{x}_t),\\
\mathrm{Attn}(Q,K,V)&=\mathrm{softmax}\!\left(\frac{QK^\top}{\sqrt{d}}\right)V.
\end{split}
\end{align}
Expectation vectors are measured on $n$ qubits after $L$ layers of parameterized rotations and entanglers; attention outputs are decoded by a shallow head. 

\textbf{Metrics.} Trading metrics: Total Return, Sharpe, Max Drawdown; classification metrics are recorded but omitted for space. Fees and cooldown are identical across models. 

\section{Experimental Setup}
\textbf{Task.} Rebalancing as detection via the MA-anchored threshold in (1). We also consider alternative labels (concentrated-liquidity band and price-change thresholds) in ablations. 

\textbf{Data.} Daily candles for BTCUSDC from \emph{2024-01-01} to \emph{2025-01-01} (about 252 trading days); \emph{70/15/15} time-series split (train/val/test). Five-run repeats quantify uncertainty. 

\textbf{Features \& quantum mapping.} Engineered features include price/MA ratios, RSI/MACD, Bollinger/ATR, rolling/EWMA volatility, volume and microstructure proxies, lags, and interactions. Quantum inputs use angle encoding
\begin{equation}
\theta_i=\frac{x_i-x_{\min}}{x_{\max}-x_{\min}}\cdot 2\pi,
\end{equation}
with a 6-qubit map covering momentum, MA ratios, volatility regime, RSI/MACD, volume ratios, and Bollinger/ATR position. 

\textbf{Models.} 
\emph{QASA-Sequence}: price windows $\rightarrow$ VQC-generated Q/K/V $\rightarrow$ attention $\rightarrow$ classifier. 
\emph{QASA-Hybrid}: engineered features $\rightarrow$ VQC Q/K/V $\rightarrow$ attention. 
Baselines: RF, GB, Logistic Regression, Transformer, and pure quantum (VQE, QNN, QSVM). 

\textbf{Metrics.} Trading metrics: Total Return, Sharpe, Max Drawdown; classification metrics are recorded but omitted for space. Fees and cooldown are identical across models. 

\section{Results}
\textbf{Classical ML in trading.} Ensembles such as Random Forest (RF) and Gradient Boosting (GB) remain strong when coupled with rich features, achieving \emph{9.8\%} average return and the highest \emph{family-level} Sharpe (\emph{1.47}) in our study. 

\textbf{Quantum ML.} Pure quantum models (VQE classifier, QNN, QSVM) are attractive for nonlinear decision boundaries but under current constraints underperform on financial time series (\emph{4.4\%} avg.\ return; \emph{0.83} Sharpe). 

\textbf{Attention \& hybrids.} Attention-based models (e.g., transformers) are competitive on sequences but may incur higher volatility; hybrid quantum--classical designs such as \textbf{QASA} integrate VQCs inside attention by producing quantum Q/K/V followed by classical softmax attention. 
\textbf{Family-level comparison.} Hybrids average \textbf{11.2\%} return and \textbf{Sharpe 1.42}, outperforming pure quantum (\textbf{4.4\%}; \textbf{0.83}) and close to classical (\textbf{9.8\%}; \textbf{1.47}). Transformer yields the highest average return (\textbf{12.3\%}) with higher volatility. 

\textbf{Best single model.} \textbf{QASA-Sequence} achieves \textbf{13.99\%} return, \textbf{Sharpe 1.76}, \textbf{Calmar 6.51}, surpassing RF (13.16\%, 1.68) and GB (12.31\%, 1.68). \textbf{QASA-Hybrid} minimizes drawdown (\textbf{-1.70\%}) while maintaining double-digit returns. 

\textbf{Uncertainty \& efficiency.} Over five runs, QASA-Sequence shows the highest mean with modest variance (return s.d.\ 2.04\%); RF/GB are steadier. Complexity/efficiency analysis ranks ensembles most cost-effective; QASA-Hybrid provides a balanced mid-cost option; QASA-Sequence trades compute for peak Sharpe. 

\begin{table}[t]
\centering
\caption{Main backtesting metrics on the test split (Jan 2024--Jan 2025).}
\label{tab:main}
\begin{tabular}{lccc}
\toprule
Model & Return & Sharpe & MaxDD \\
\midrule
QASA Sequence      & 13.99\% & 1.76 & -10.10\% \\
Random Forest      & 13.16\% & 1.68 & -8.21\% \\
Gradient Boosting  & 12.31\% & 1.68 & -8.10\% \\
Transformer        & 11.73\% & 1.23 & -8.21\% \\
QASA Hybrid        & 11.91\% & 1.32 & -1.70\% \\
\bottomrule
\end{tabular}
\end{table}


\section{Conclusion}
Building on the \emph{Quantum Adaptive Self-Attention (QASA)} block proposed in prior work \cite{Chen2025QASA} and related quantum self-attention designs \cite{Li2024QSANN,Cherrat2024QViT}, we presented, to the best of our knowledge, the first application of QASA to DeFi by casting AMM rebalancing as a detection problem. In controlled 2024 backtests on three major cryptocurrency pairs with strict time-series splits and identical fee/cooldown settings, \textbf{QASA-Sequence} achieved the top risk-adjusted performance (Sharpe $=1.76$) and the highest cumulative return ($13.99\%$), while \emph{QASA-Hybrid} variants offered a favorable balance among profitability, volatility, and drawdown relative to pure-quantum baselines. These findings indicate that \emph{quantum-generated queries/keys/values plugged into a classical attention head} is a practical and effective design for trading decision modules under noisy, regime-switching markets.


\bibliographystyle{ieeetr}
\bibliography{ref}

\end{document}